# MetaMetaZipf. What do analyses of city size distributions have in common?

**Clementine Cottineau, CNRS, August 2020**

## Abstract.

In this article, we conduct a textual and contextual analysis of the empirical literature on Zipf's law for cities. Building on previous meta-analysis material openly available, we collect full texts and bibliographies of 66 scientific articles published in English and construct similarity networks of the terms they use as well as of the references and disciplines they cite. We use these networks as explanatory variables in a model of the similarity network of the distribution of Zipf estimates reported in the 66 articles. We find that the proximity in words frequently used by authors correlates positively with their tendency to report similar values and dispersion of Zipf estimates. The reference framework of articles also plays a role, as articles which cite similar references tend to report similar average values of Zipf estimates. As a complement to previous meta-analyses, the present approach sheds light on the scientific text and context mobilized to report on city size distributions. It allows to identified gaps in the corpus and potentially overlooked articles.

## 1. Introduction

The parallel development of more data accessible at city level on the one hand and of non linear regularities being found as a marker of complexity (in networks, in organisms, in cities) have produced a regain of interest in the study of city size distributions in recent years, and more specifically a renewal of the appraisal of Zipf's law for cities. Of particular importance in these debates are the definition of the objects studied (i.e. the limits, thresholds and components of cities which affect the population included or not), the model to summarize the distribution (power-law, lognormal, polynomial) and its fit to the data (fitting procedure, value of the power exponent, uncertainty). It is usually agreed upon that city populations follow a heavy-tail distribution in most countries or regions and at most time periods, although the precise form of the distribution and the estimation of its main parameters tend to vary. The universality claim of Zipf's law (1949) can thus be accepted with respect to the general trend, but is rejected in its strictest form (i.e. a power law of exponent -1 between city sizes and their ranks by size). Previous meta-analyses of studies providing an empirical estimation of Zipf's exponent have shown indeed that on average, empirical estimations tend to deviate from the strict value of -1 (Nitsch, 2005; Cottineau, 2017)[1]. A share of such deviations can be attributed differences in the technical specifications of the studies (their total number of estimates, the range of countries and periods analysed) and of the empirical estimation (delineation of cities, thresholds, estimation procedure, etc.). A smaller share of the variance can be attributed to territorial characteristics of the city system (its level of urbanisation for instance) and no share has been found to vary significantly with planning actions (Cottineau, 2017). Therefore, empirical deviations to Zipf's law remain for the most part unexplained or unexplored. Publication biases as well as differences in reference frameworks and disciplinary traditions might generate systematic differences in the measuring and reporting of empirical distributions of city sizes, but they are unobservable with a traditional meta analysis. For example, despite addressing the same empirical estimation of Zipf's law (same country, same set of city, same date, same estimation method), there can be strong differences in the way the papers from the meta analysis frame, exploit and report on this result, depending on the aim of their research ("proving that Zipf's law is a universal feature of urban systesm", "showing that the lognormal form is better suited", "looking for national differences in urban hierarchy", etc.). The empirical results of such studies could then appear clustered by different school of thoughts. The present work therefore asks the question: what do analyses of city size distributions have in common? It goes a step further in the secondary

1 In the rest of the paper, we consider the absolute value of the exponent, called "alpha" and its deviation from 1.

analysis of Zipf's law for cities, by exploiting similarity networks drawn by the studies included in a meta analysis. Building on the open-source corpus of MetaZipf (Cottineau, 2017), which contains 1962 empirical estimations of Zipf's exponent alongside their technical and territorial specifications from 86 studies, it characterises the pairwise similarities of studies based on their bibliographies, their textual content and their disciplinary exposure[2]. Combined with the pairwise similarities of the study content, it aims to reveal new insights about the deviation of estimated exponents in their published results. We find evidence that pairs of articles with similar wording and similar bibliographies tend to report similar average values of estimates. Similar wording also correlates positively with a similarity in the level of dispersion of values reported. The data and code of the present study has been made fully open on github, including an R notebook with all visualisations[3].

## 2. Why a "meta"-meta-analysis?

Meta-analyses are important tools to reflect on the collective production of an established field of inquiry, especially when it produces quantitative estimations and prediction statements. In that respect, city size distributions and their modelling with power laws date back more than a century (Auerbach, 2013), and still generate dozens of dedicated articles every year. Such scientific productions originate from a diversity of disciplines and research domains such as economics, geography, statistics, physics, regional science, planning and mathematics. Consequently, authors of studies included in a Zipf meta-analysis tend to publish in a diversity of journals which all have different formal and theoretical requirements: the size of text, the type of proofs received as valid, different evaluations of the necessary, legitimate and superfluous references. For instance, economics journal usually require econometric models with controls and specific way of presenting results in standardised tables. Physics journals tend to publish shorter articles with large supplementary materials. "*Planning papers tend to cite eclectically. [...] This will be a feature of social science in general compared with science journals but, within the social sciences, one might expect certain broader applied subjects such as planning to be especially unfocused in the literature they cite. […] Planning papers are also eclectic in the type of references cited reports and plans as well as academic papers and this may lower impact statistics.*" (Webster, 2006, p.488). Journals in geography will tend to favour analyses of spatial variations of a given phenomenon while other subjects will look for its regularity. Could such "meta" properties also signal differences in the definition of the aim of the research, the design of the experiment and eventually the value of the reported results, thus offering a new angle to explain their difference? The hypothesis leading this research is that they could. Indeed, science is a social practice performed by actors embedded within institutions, disciplinary frameworks and legacies (Latour, 1986). It would therefore be possible to suggest that studies written in a similar way, citing similar references and publishing in the same kind of journal exhibit more similar results (controlling for the object of their study, in our case: the similarity of cities, countries and time periods studied) than studies which originate from very different fields, point to very different bibliographies and use distinct scientific languages. There is evidence from the MetaZipf corpus that a significant diversity of languages, reference frameworks and disciplines exist. For example, Gabaix & Ibragimov's (2011) article is built like a mathematical demonstration (using terms like "theorem" and "lemma" several times) whereas other articles read more like monographies. Some articles systematically reference back to Zipf (1949) and Auerbach (1913), whereas others start the debate where Gabaix (1999) left it. Some articles cite a very large number of external references (Parr, 1985 or Berry & Okulicz-Kozaryn, 2012) when others do not (such as articles published early or in physics journals). Finally, the range of journals cited and chosen for publication is broad, it ranges from mainstream economics to specialised geography, statistical physics and beyond. The objective of the present work is to assess whether such diversity is reflected systematically in the variation of results reported, in order to better understand urban hierarchies around the world (rather than the scholars who study them).

2 Elements which, incidentally, loop us back to Zipf's original research in linguistics.

3 http://clementinecttn.github.io/MetaZipf/metametazipf_notebook.nb.html

## 3. Methods and materials

This section details the collection of the "meta"-meta-data and the strategy used to convert it into pairwise similarity matrices along a number of dimensions. It also presents the model used to regress differences in the reported distribution of Zipf estimates by meta-properties, controlling for technical and territorial specifications. The material of the present study consists in a corpus of studies which have all published estimations of Zipf's power law exponent on empirical city size distributions. It makes use of the openly available database MetaZipf[4], which contains 1962 such empirical estimations from 86 studies, along with their specifications. The 86 studies have been selected to fulfill three criteria: "*they contain at least one estimate of the rank-size exponent based on population ; the regression is made on empirical urban data; the regression model is bivariate (i.e. relating populations and ranks or ranks—1/2, but not to any other instrumental variable).*" (Cottineau, 2017, p. 4). In the present work, only 66 of them fulfilled additional criteria detailed below. This subset of 66 studies are subsequently referred to as "the corpus".

### 3.1. Collecting full-texts

For an article from the MetaZipf database to be included in "the corpus", it has to available in open-access or accessible with an extensive institutional subscription, in a machine-readable format. Additionally, only published journal articles written in English were selected, in order to run a coherent textual analysis. This excludes texts in other languages and formats, such as books and dissertations. This choice is detrimental to the recognition that science is plural in forms, languages and origins. However, it did not affect the original sample too much, since most references in MetaZipf were already predominantly in English and in an article format. The corpus is thus composed of 66 full-texts of English-written articles. Out of the original document, only the body of the text was retained. This means that titles, affiliations, abstracts, keywords, section titles, figures, tables, equations, references, footnotes and line breaks were removed. The remaining text was used for text mining analysis, after a traditional automated treatment (with the R 'tm' package, cf. Feinerer et al., 2019) in order to remove punctuation, numbers and stop-words and transform the remaining word to lower case. Term frequencies were attached to each reference to allow for a study of wording similarity between them.

After treatment, corpus articles exhibit a continuous array of sizes (figure 1A), from 384 words for Popov (1974) to 5522 words for Ignazzi (2014). Apart from significantly shorter sizes in physics articles (around 1600 words on average per corpus article, compared to 3000 on average in economics and 2500 in geography), we do not find any trend by year of publication or else.

### 3.2. Collecting citations

To explore the citation network of corpus articles, each reference from the 66 english-written articles was recorded and formatted in a way that allows to query the authors' names, the year and journal of publication. The 66 corpus articles generated 304 internal citations (i.e. to other articles included in the corpus) and citations to 1155 distinct external references (including references to articles, reports, books or dissertations in various languages) from over 700 different journals or publishing institutions. Corpus articles exhibit once again a disparity of (external) bibliography sizes (figure 1B), from 6 items in Suarez-Villa (1980) and Popov (1974) to 76 in Berry & Okulicz-Kozaryn (2012). Apart from significantly shorter sizes in physics articles (around 15 items on average, compared to 22 on average in economics and 24 in geography and regional science), we do not find any trend by year of publication or else.

The journals most frequently chosen to publish corpus articles (figure 2A) coincide with the journals where bibliographical references most frequently come from (figure 2B), i.e. Urban Studies and the Journal of Regional Science, the Journal of Urban Economics, Regional Science and Urban Economics or the Journal of Economic Geography. The average year of publication in the corpus is 2004, ±1 year for articles from different disciplines except articles published in physics journals,

4 https://github.com/ClementineCttn/MetaZipf

whose interest in city size distributions and average year of publication is more recent (2013). By contrast, the average year of publication for external references is 1989.

**Figure 1. A (left). Distribution of text size in the corpus (number of non-stop-words). B (right). Distribution of bibliography size in the corpus (number of external references).**

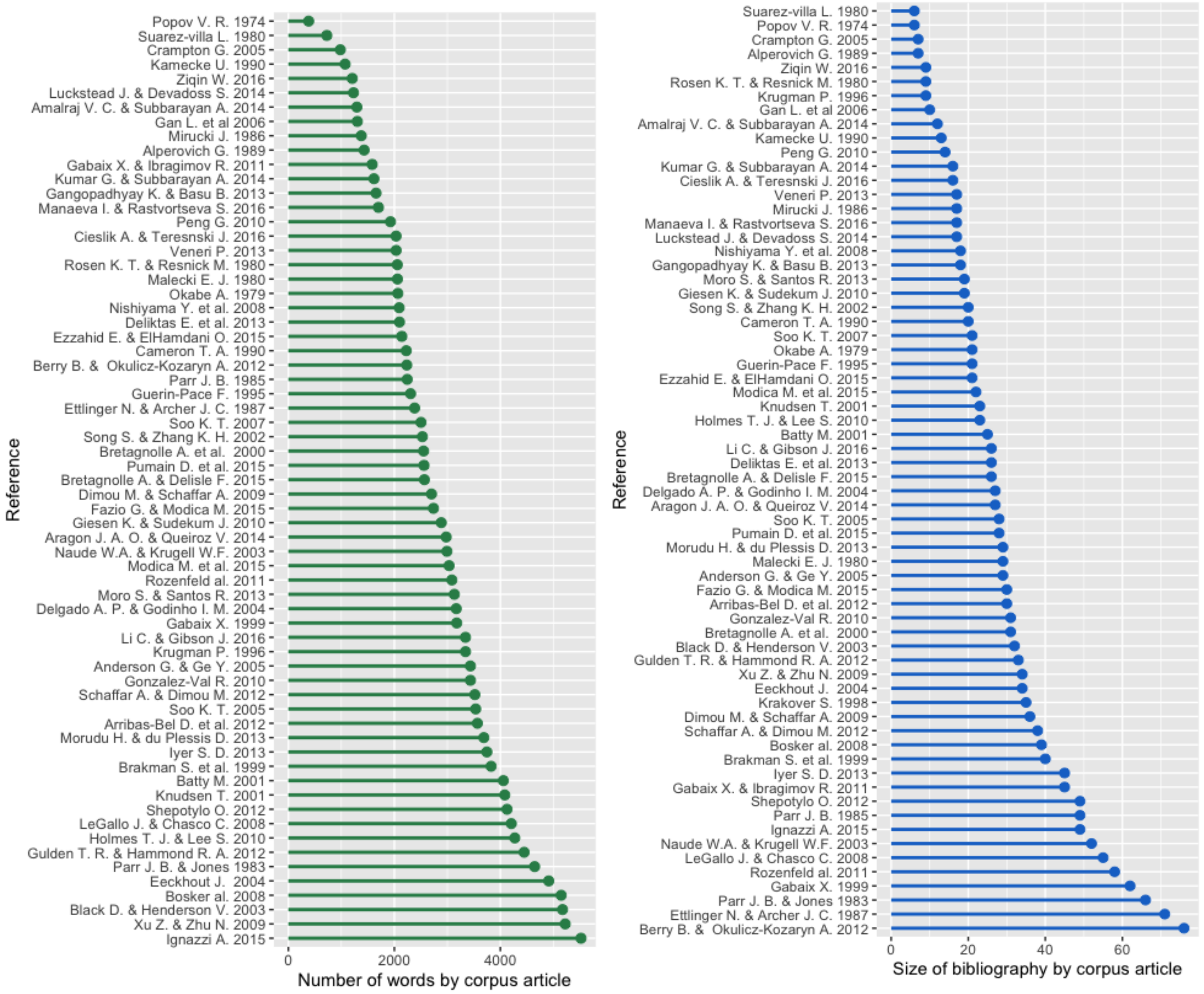

The most cited external reference is to Zipf himself. 37 out of the 66 corpus articles cite it for its 1949 book on the "principle of the least effort" and 5 cite it for its 1941 work "National unity and disunity; the nation as a bio-social organism". Corpus papers not citing any of the two Zipf references are frequent among those published at earlier dates (figure S1 in Supplement), and proportionally more in geography and economics journals where it might be considered obvious. By contrat, 4 out of 4 articles published in physics journals cite Zipf. It is interesting to note, however, that Zipf's work is not the most cited reference in the corpus: two internal references appear even more frequently (figure 3A): Gabaix's theoretical 1999 paper (41 out of the 50 other articles published in or after 1999) and Rosen & Resnick's comparative 1980 paper (39 out of the 62 other articles published in or after 1980). Externally (figure 3B), the reference to Auerbach's work from 1913 is in the top 3 with 23 external references, but it is less frequently cited than Gabaix & Ioannides's (2004) chapter in the Handbook of Urban and Regional economics, with external citations from 29 corpus articles.

**Figure 2. A (left). Distribution of corpus articles by journals (and series) publishing them. B (right). Distribution of articles cited by corpus articles by journals (and series) publishing at least 5 of them.**

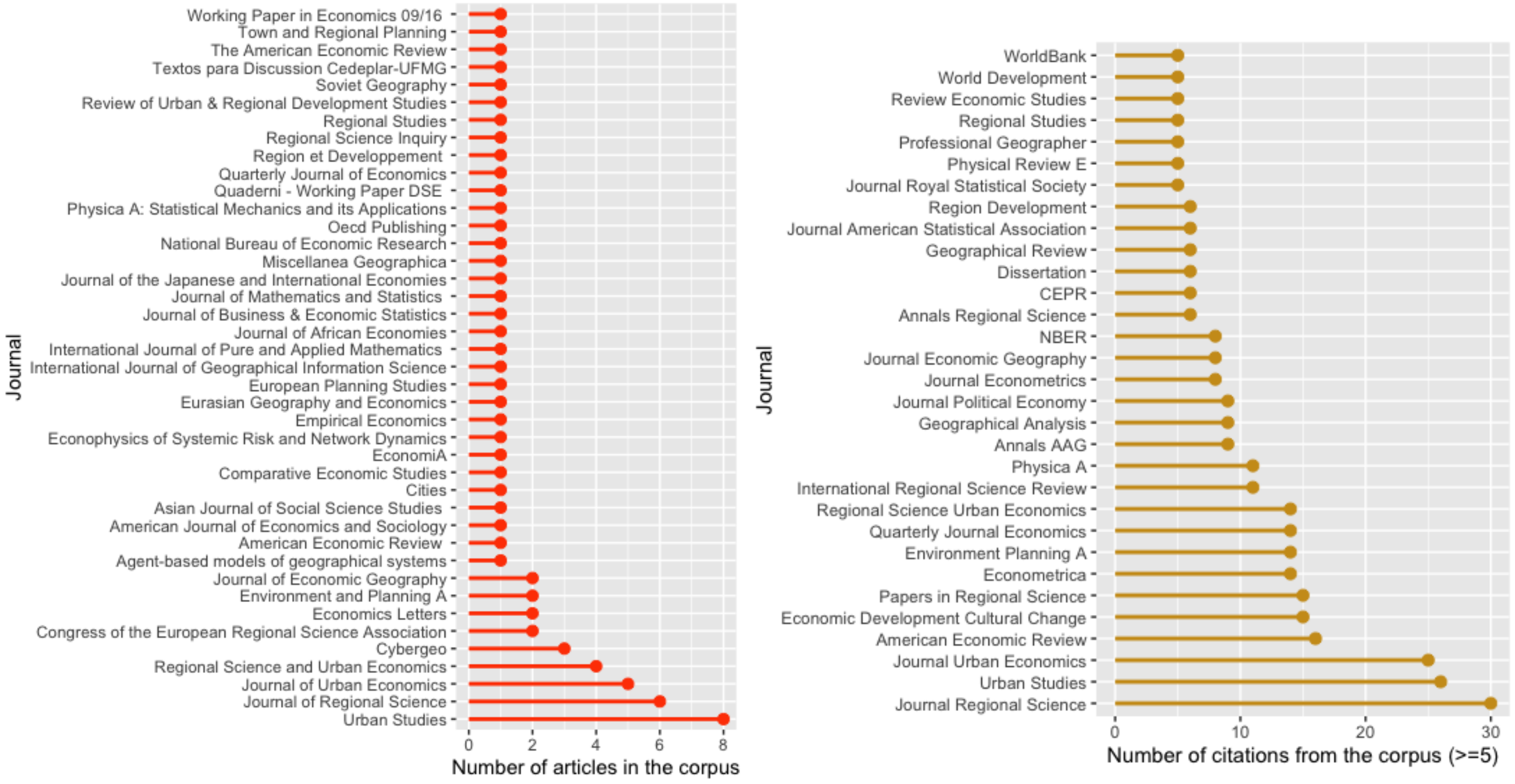

The graph on figure 3B shows that many top references cited externally are early classics of urban theory (Christaller, 1933 [6 cites], Losch, 1940 [8]) and statistics (Gibrat, 1931 [16 cites], Simon, 1955 [14], Pareto, 1897 [8], Hill, 1975 [7]). Some highly cited references such as Singer, 1936 [11] or Eaton & Eckstein, 1997 [23] actually include empirical estimations of Zipf's exponent, which suggests that they could have been included in the corpus. However, the former was not accessible and the latter contains instruments in the regression. It could be considered to relax this criterion to include its findings in the future, given its influence on the corpus' reference frameworks.

The most striking feature of this list however is the prominence of post-1995 contributions from three economists in the top cited references (Gabaix, Krugman and Ioannides) compared to earlier works by geographers (like Berry in 1961, Parr since the 1970s, or Moriconi-Ebrard in the early 1990s). As pointed by C. Webster (2006, p. 489-90) in the context of planning journals, "*there is both a publishing and a cognitive limitation on the number of citations included in a paper and this means that the rate of citation growth will be higher, the higher the citation count of a paper. Well-cited papers will become more well cited. If the total number of citations per paper grew to accommodate the increasing number of papers as a field grows, then this inequality might not be inevitable. But reference lists do not get ever longer and, as a result, the frequency of paper citation counts tends to follow a rank-size pattern (sic)*". In the case of top cited papers in this study, they indeed belong to highly visible academics of large, established and dominant disciplinary fields, whose articles in general and the Zipf ones in particular, generate hundreds to thousands of citations (2133 for Gabaix's 1999 "Zipf's law for cities, an explanation"). Finally Nitsch (2005)'s meta analysis is frequently cited (21% of all corpus articles and 34% of those published in or after 2005). Many externally cited references do not appear on this graph for they receive less than 5 mentions from the 66 corpus bibliographies[5].

---

5 Some of them are indeed quite specific, for instance those from the aerosol literature cited in to Eeckout (2004): **Haaf, Amin and Jaenicke, Rainer.** "Results of Improved Size Distribution Measurement in the Aitken Range of Atmospheric Aerosols." *Journal of Aerosol Science*, 1980, *11*(3), pp. 321–30. & **Hinds, William C.** *Aerosol technology*. New York: Wiley, 1982.

**Figure 3. A (left). Distribution of citations to corpus articles by corpus articles. B (right). Distribution of citations (over 5) to non-corpus articles by corpus articles**

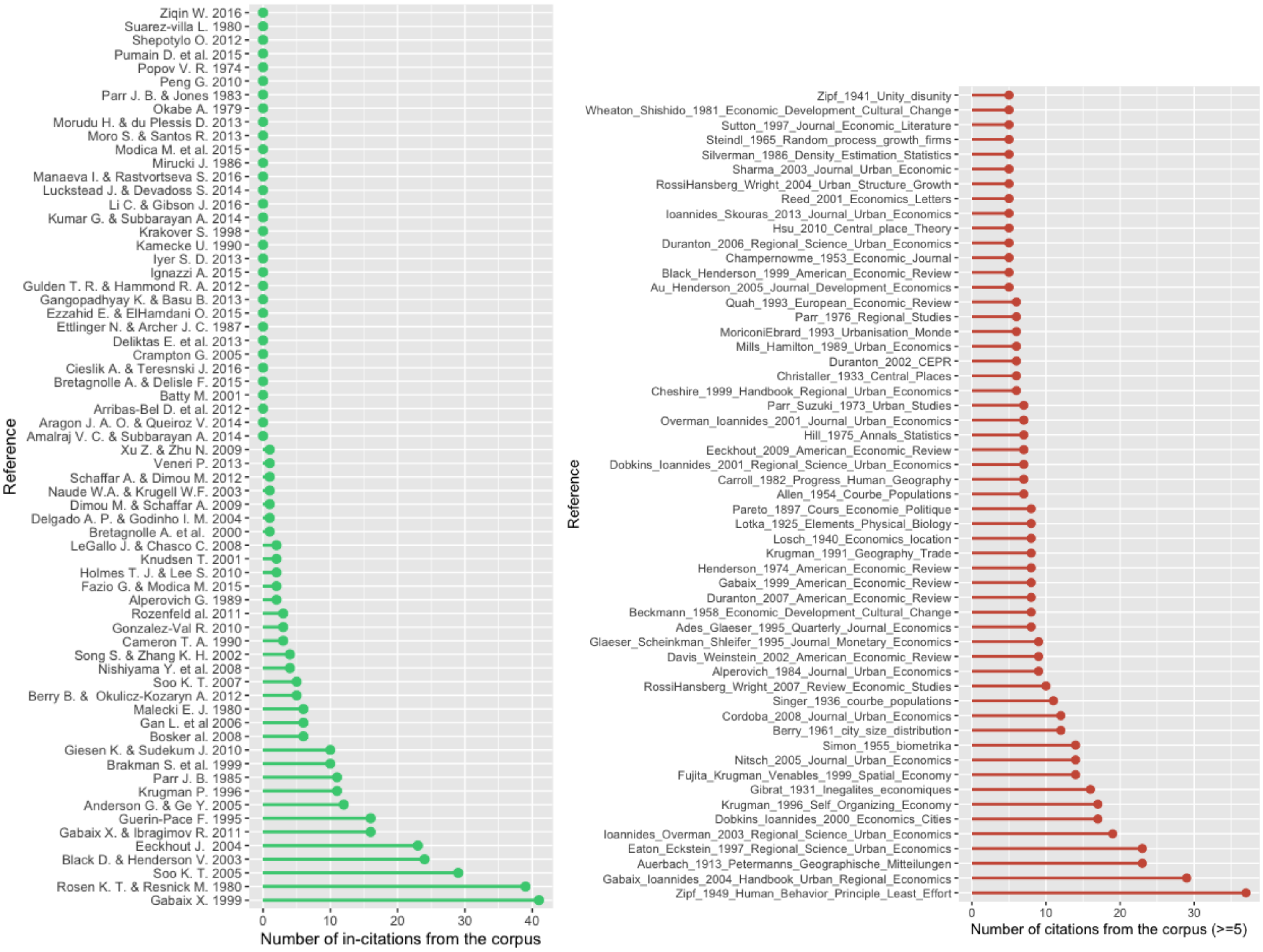

### 3.3. Translating journals into disciplines/disciplinary fields

In order to study the disciplinary dynamics of corpus articles on Zipf's law for cities, we assigned a "discipline" to each of the 707 journals and publishing institutions from which internal and external references of this meta-meta-analysis were taken. We identified 5 fields: Economics (ECO), Geography (GEO), Regional Science and Planning (REG), Statistics (STAT) and Physics (PHY). Although identification of journals in the last two fields was rather straightforward, the lines between Economics, Geography and Regional Science were quite blurry. However, it seemed interesting to distinguish them for two reasons. Firstly, economics and geography are recognised disciplines whose practitioners do not frequently publish in each other's journals, whereas Regional Science sits precisely at the intersection between economics, geography and planning. In regional sciences/studies conferences and journals, it is not unusual to find references to both disciplines. We thus wanted to see if regional science occupied this middle ground position in city size distribution studies as well. Secondly, the separation acknowledges the fact that publication strategies vary greatly between the journals of these fields, in terms of exposure, sphere of impact, formal and theoretical requirements, even when articles deal with the same object.

The keys used to allocate journals between the three fields were the following:

- "ECO" for general economics journals (such as the Quarterly Journal of Economics) as well as journals with "economics" in their names (Journal of Urban Economics for instance), if they do not have "regional science" in it as well.

- "REG" when the subject is "urban affairs", "urban studies", or has "urban and regional" in the name.
- "GEO" for general geography journals (for example, the Annals of the Association American Geographers) as well as journals with names referring to the processes of urban development and urbanisation.

This approach contains some ad-hoc character. We have tried to alleviate it first by providing access to the lookup table. We are aware of existing journal classifications but find that they do not reflect entirely the stakes of this sub-field (nor do they provide guidance for the classification of books, reports and dissertation). Second, an assessment of the most frequent journals with the Scimago classification shows for instance that the ad-hoc fields we attributed to journals matches at least one of the Scopus subject areas proposed by Scimago[6], considering that "Environmental Science (miscellaneous)" corresponds to Regional Science and planning (table S1 in supplement). The advantage of our system is that it provides a single category for each source, whereas Scimago has a varying number of entries for different journals, and no entry for French journals like "Région et Développement" which is externally cited 6 times in our corpus, or for dissertations and World bank databases.

After applying this ad-hoc translation to all external references, we can see a stark difference between the distribution of corpus articles by disciplinary field and that of their bibliography (table 1). Indeed, while most corpus are published in regional science and economics journals, but their framework of reference comes primarily from economics and geography, or at least from articles published in economics and geography journals. Secondarily, corpus articles draw from the statistics (for estimation methods and tools) and regional science literatures. Thirdly, they cite articles published in physical science journals. The large number of "OTHER" references indicates the diversity of Zipf-related work bibliographies, which frequently reference other disciplines (political science, architecture, etc.) and formats (reports, dissertation, etc.).

**Table 1. Distribution of references by discipline of the journal they were published in.**

| **Disciplinary field** | **ECO** | **GEO** | **STAT** | **REG** | **PHY** | **OTHER** |
|---|---|---|---|---|---|---|
| **Corpus** | 23 | 13 | 0 | 26 | 4 | 0 |
| **External references** | 341 | 233 | 126 | 125 | 49 | 281 |

### 3.4. From individual studies to reference networks

In order to assess whether the common meta properties of articles dedicated to the empirical estimation of Zipf's law play a role in the variation of results they report, we constructed nine non-oriented networks of similarity between corpus articles. The networks all have 66 vertices corresponding to corpus articles. They differ in the distribution of edge weights connecting vertices. The first three networks ("wording", "citation" and "discipline") were built to test our three hypotheses: that similarity in text, citation and discipline contexts signal similarity of goals and research design resulting in similarities in values reported for Zipf's exponent alpha. The next three network ("country", "decades" and "city definition") were built to control for the similarity in the objects actually studied by corpus articles. The networks "mean alpha" and "sd alpha" are the one to enventually "explain": they correspond to the networks of corpus articles drawn by the similarity of the distribution of Zipf estimates they report (with the mean and standard deviation - sd). A last network ("n alpha") draws the similarity of corpus articles based on the number of estimates they report, which has a direct influence on standard deviation calculation and is thus used as a control variable. Similarity for all networks was measured pairwise, using the cosine similarity (if not stated otherwise). A subset of each network is visualised using the 'igraph' R package (Csardi

6 https://www.scimagojr.com/ (accessed on 20/08/2020)

& Nepusz, 2006). For better visibility, we apply a cutoff to the weight of edges represented, exclude non-connected vertices, and colour vertices using Louvain community clusters (except in figure 6). These representations provide clues for interpretation. However, the modelling analysis is run on the complete networks. The entire analysis is available and reproducible online[7].

*3.4.1. The "wording" network.*

The "wording" network represents the similarity between corpus articles based on the frequency of words they have used to write their paper and present empirical estimations of Zipf's exponents. Using the 66 full texts collected, we computed the frequency distribution of 10,791 non-stop words in each articles. The vectors used as inputs for the "wording" cosine similarity are thus composed of 10,791 values comprised between 0 and 1. A visualisation of a subset of the network created is visible on figure 4, along with some of the most frequent terms used by corpus articles of the different communities. The variation is vertices sizes represents their total number of terms.

**Figure 4. Similarity network of corpus articles by frequency of words used (cut-off at 0.65).**

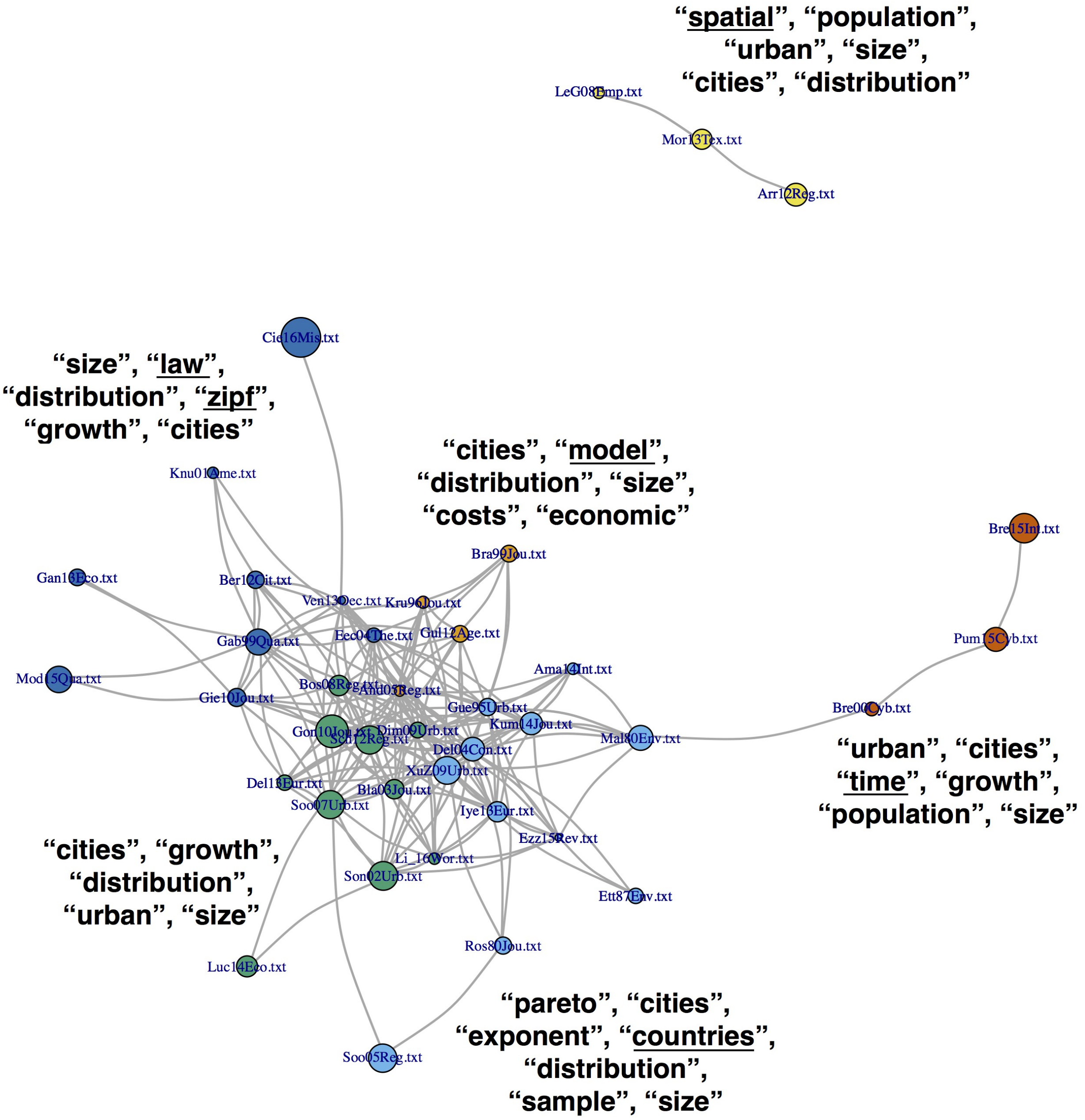

** cf. Table S0 in supplement for a lookup table between article identifiers and bibliographic references.*

---

7 https://clementinecttn.github.io/MetaZipf/metametazipf_notebook.nb.html

The figure shows a network with strong connectivity. Indeed, most articles use, at the very least, the words “cities”, “size” and “distribution” very frequently. However, a disconnected community of three articles (Le Gallo & Chasco 2008; Moro & Santos, 2013 et Arribas-Bel et al., 2012, in yellow) exhibits a more important use of the word “spatial”. These works even have “spatial” in their title. Their aim is not to verify Zipf's law but to present and analyse a national urban system, respectively Spain, Brasil and Australia. Another cluster (in red) shows a specific use of the term “time”. Originating from a single research group in France, Bretagnolle et al. (2000, 2015) and Pumain et al. (2015) indeed present long-term evolutions of systems of cities, reporting on their growth and structure of several decades. The light blue cluster gathers comparative studies who therefore make a more thourough use of the term “countries”. Two other clusters represent articles less devoted to empirical analysis and more to the testing of “Zipf”'s “law” (dark blue) or “model”-ing its generation (orange). This network thus represent the way Zipf's law is approached by authors and signal somehow the finality of the argument and how estimation results will be used.

*3.4.2. The “citation” network.*

The “citation” network represents the similarity between corpus articles based on the external references they cite in bibliography. It could be argued that two papers citing the exact same corpus of references would more frequently share the same aim, such as “proving” or “disproving Zipf's law”, and therefore report more similar estimate values. The similarity was measured from the 66 vectors of 1155 external references, coded 1 if the reference was cited and 0 otherwise. A subset of the network is visible in figure 5, with the size of vertices showing the total number of citations.

**Figure 5. Similarity network of corpus articles by external articles cited (cut-off at 0.25).**

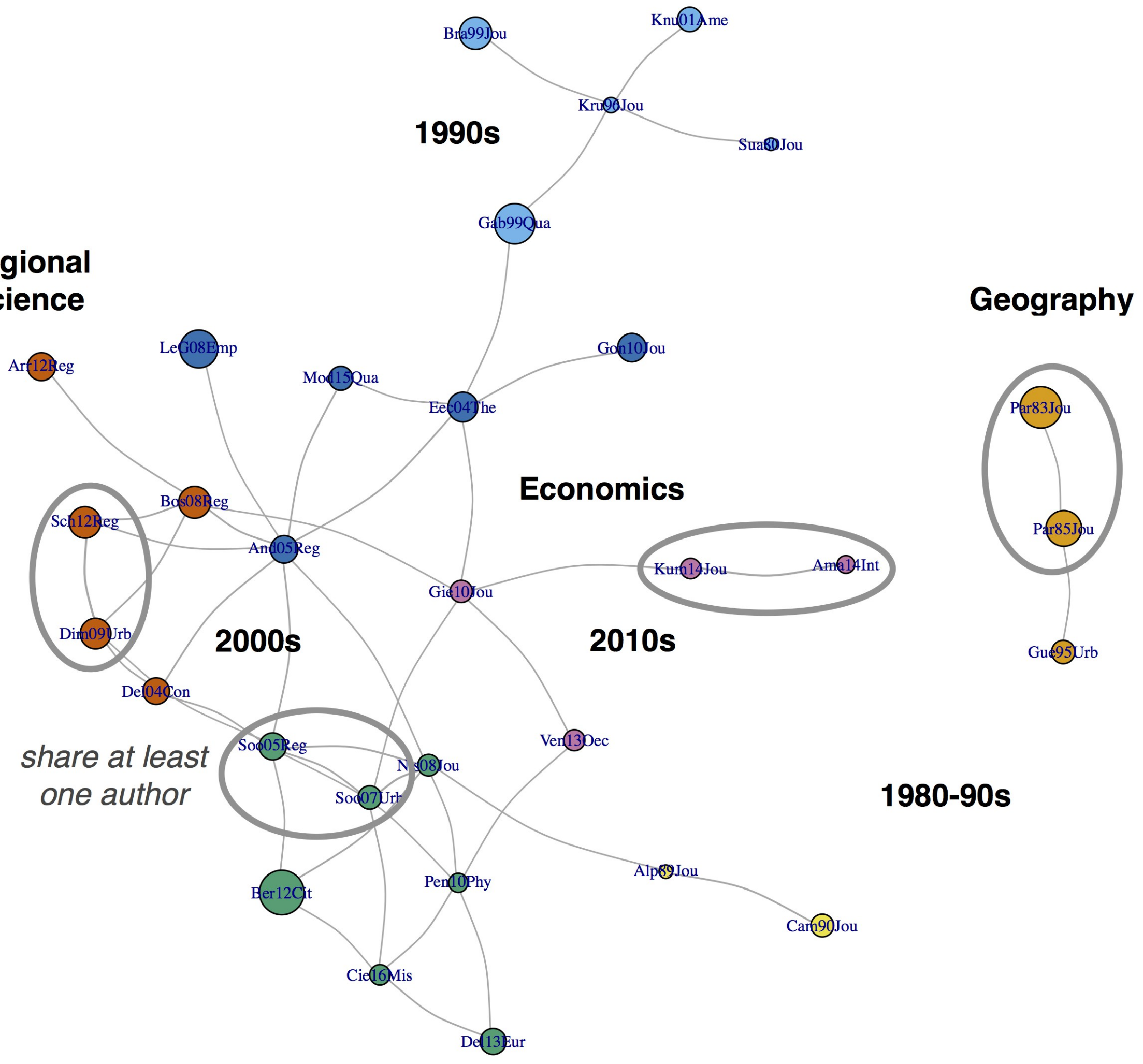

This network is less connected than others, suggesting that the reference framework of authors depends on a diversity of factors besides a common object of study. Indeed, the subnetwork shown in figure 5 is organised along time, co-authorship and disciplinary lines. The similarity of citations is, quite trivially, lower for articles of different periods because of the unavailability of later references for earlier articles. Therefore, we see clusters of corpus articles with similar publication dates (orange and yellow clusters in the 1980s and early 1990s, light blue cluster in the late 1990s, red and green clusters in the 2000s and 2010s, pink cluster in the 2010s). The strong similarity of co-authored articles (Soo, 2005 and 2007 or Dimou & Schaffar, 2009 and Schaffar & Dimou, 2012 for instance) reveal the inertia of reference frameworks over time and their relative subjectivity. Finally, articles published in economics journals seem to share more bibliography between them than they do with articles published in geography.

*3.4.3. The "discipline" network.*

The "discipline" network represents the similarity between corpus articles based on the discipline of the journal their external references were published in. The cosine similarity was measured on vectors of 6 items (the number of external refences from each discipline). For this representation, we did not use community clusters for colouring nodes but instead the discipline of the journal where corpus articles were published.

**Figure 6. Similarity network of corpus articles by external disciplines cited (cut-off at 0.9).**

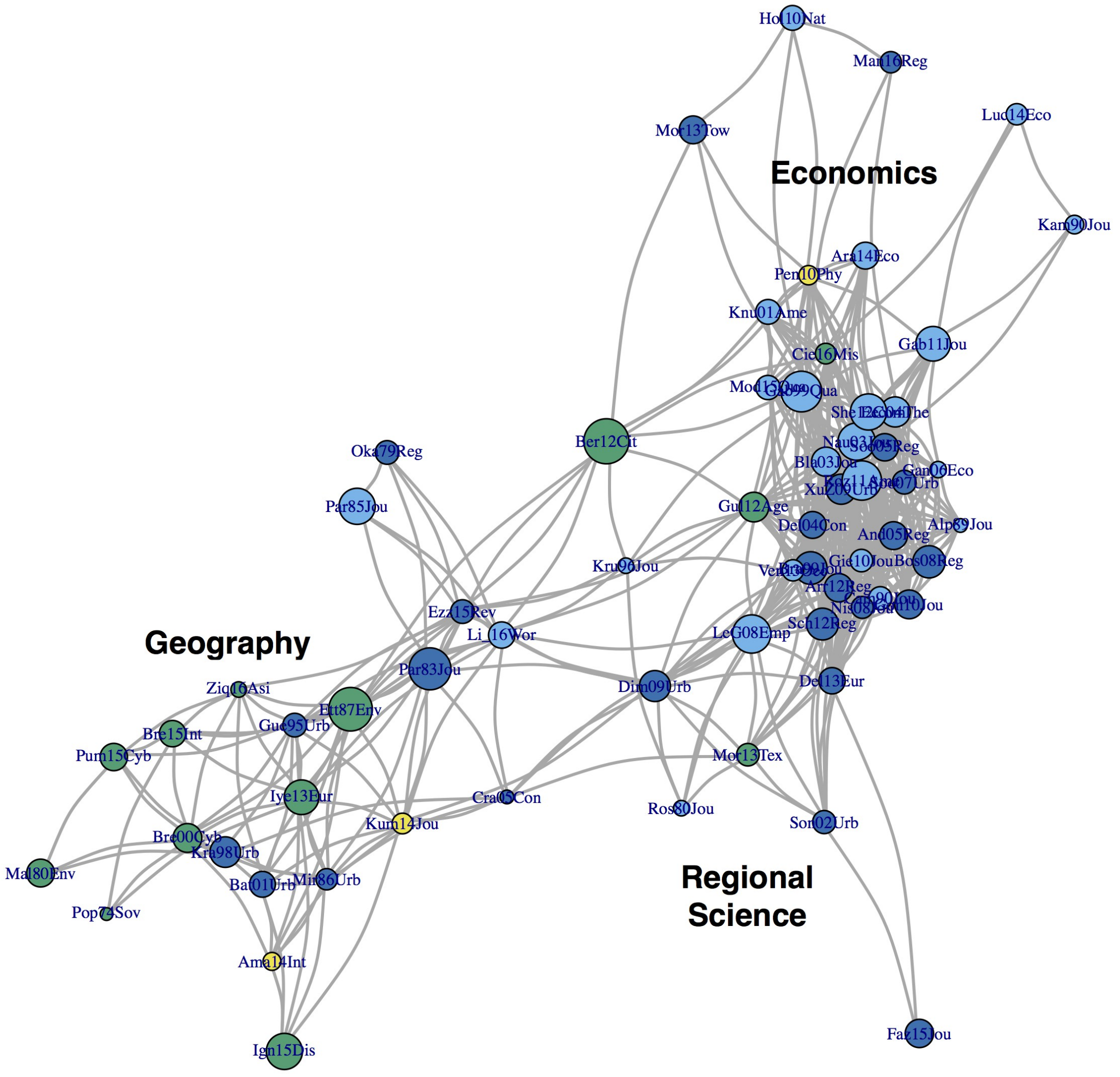

**Node colours show the discipline of the article rather than its membership to a Louvain community cluster. Yellow: physics. Green: geography. Light blue: economics. Dark blue: regional science and planning.*

The figure shows that some entanglement of disciplinary references, with some regional science corpus articles citing a similar pool of disciplines as geography corpus articles. However, this might be an artifact of publication strategies, since the articles in question (such as Parr & Jones, 1983, Guérin-Pace, 1995 or Batty, 2001) are authored by people also recognised as geographers. Corpus articles in economics and in regional science also share similar disciplinary references. The divergence of disciplinary framework appears main between geography and economics articles, although some articles (Krugman, 1996; Berry and Okulicz-Kozalyn, 2012 or Dimou & Schaffar, 2009) work as bridges, citing from a more varied pool of disciplinary references.

*3.4.4. The "country" network.*

The "country" network represents the similarity between corpus articles based on the countries on which they perform empirical estimations of Zipf's exponents (figure S2). High similarity characterise articles dedicated to the same area (USA, China, South Africa) and articles dedicated to comparative studies (like the most extensive of that kind: Soo, 2005; Rosen & Resnick, 1980). The two densest clusters are composed by studies reporting Zipf's estimates exclusively for American (in orange) and Chinese (in blue) cities.

*3.4.5. The "decades" network.*

The "decades" network represents the similarity between corpus articles based on the decades on which they perform empirical estimations of Zipf's exponents (figure S3). High similarity characterise articles dedicated to the same period. The densest clusters are composed by corpus articles reporting Zipf's estimates exclusively for a single decade (such as Cameron, 1990 or Krugman, 1996).

*3.4.6. The "city definition" network.*

The "city definition" network represents the similarity between corpus articles based on the city definition used to collect city populations (municipalities, agglomerations or metropolitan areas mostly) on which empirical estimations of Zipf's exponents are performed (figure S4). This network is polarised by the use of one or more city definitions in the corpus article. The larger cluster (in orange) unfortunately reflects the fact that most city size distributions are analysed within improper urban delineations (the 'city proper' or municipal boundaries), as their shape and defining principles vary greatly across countries while they tend to stay fixed over time whereas cities expand spatially and functionally.

*3.4.7. The "alpha" networks.*

The "alpha" networks represents the similarity between corpus articles based on the distribution of empirical estimations of Zipf's exponents (alpha expressed under the Lotka form, or 1/alpha expressed in the Pareto form) they report. We choose to model two aspects of this distribution: the average value of alpha reported on the one hand, and its standard deviation on the other hand. Additionally, we use the number of estimates reported as an extra control.

To construct the "mean alpha" network, we compute the average value of alpha estimates $\bar{a}_i$ per study i and the average value of alpha estimates $\bar{a}$ for the entire sample (1962 estimates in total). We then compute a distance $d\bar{a}_{ij}$ between studies as follows:

$$d\bar{a}_{ij} = | \bar{a}_i - \bar{a}_j | / \bar{a} \quad , \text{with } i \neq j$$

The smaller this distance the more studies i and j report Zipf estimates close in value. To transform this distance into a similarity, we simply multiply $d\bar{a}_{ij}$ by -1. The network emerging from this similarity is therefore organised around groups of studies based on the average values of alpha estimates they report (figure 7). At the low end of the sprectrum, studies like Holmes & Lee (2010) or Kumar & Subbarayan (2014) report very low values of estimates (0.75 on average for the group in red), which indicates city sizes more evenly distributed than predicted by Zipf. At the other end of the spectrum, studies like Ziqin (2016) or Nishiyama et al. (2008) report high values of estimates (1.15 on average for the group in orange), which reflects highly uneven city size distributions.

**Figure 7. Similarity network of corpus articles by average value of estimates reported (cut-off at -0.025).**

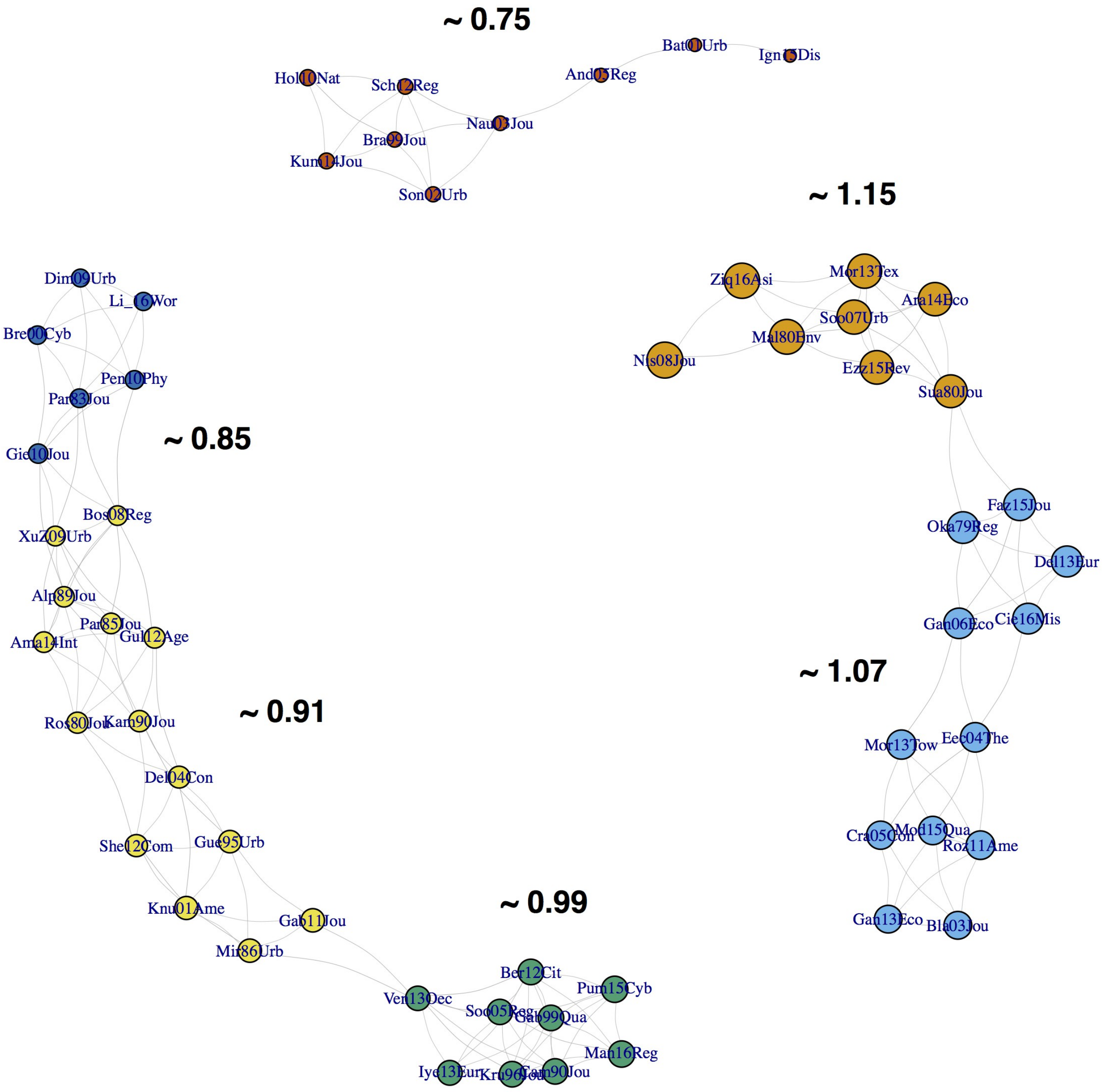

** the size of nodes reflects the average value of estimates reported in the article and the numbers in black correspond to the average value reported for the community. cf. Table S0 in supplement for a lookup table between article ID and bibliographic references.*

To construct the "standard deviation" network, we compute the standard deviation $\sigma^2_i$ of alpha estimates per study i and the standard deviation of alpha estimates $\sigma^2_a$ for the entire sample. We then compute a distance $D\sigma^2_{ij}$ between studies as follows:

$$D\sigma^2_{ij} = | \sigma^2_i - \sigma^2_j | / \sigma^2_a \quad , \text{with } i \neq j$$

The smaller this distance the more studies i and j report a similar dispersion of Zipf estimates. To transform this distance into a similarity, we simply multiply $D\sigma^2_{ij}$ by -1. The network emerging from this similarity is therefore organised around groups of studies based on the average dispersion of alpha estimates they report (figure 8). At the low end of the sprectrum, studies like Okabe (1979) or Gabaix (1999) report estimates very close to one another (0.02 standard deviation on average for the group in orange), frequently because such studies only report only 1 or 2 estimates. At the other end of the spectrum, studies like Eeckout (2004) or Fazio & Modica (2015) report very dispersed distributions of estimates (0.43 standard deviation on average for the group in dark blue). In these two examples, such dispersion is produced by estimations all made for the USA in 2000 and 2010, but with large variations of truncation points (i.e. the minimum population size to include cities in the sample), from 135 residents to 29,000, which changes the number of places included in the regression from about 156,000 to only 35. As noted in Cottineau (2017), the truncation point is one of the most important technical specifications with respect to the variation of Zipf's estimates in the literature.

**Figure 8. Similarity network of corpus articles by standard deviation of estimates reported (cut-off at -0.1).**

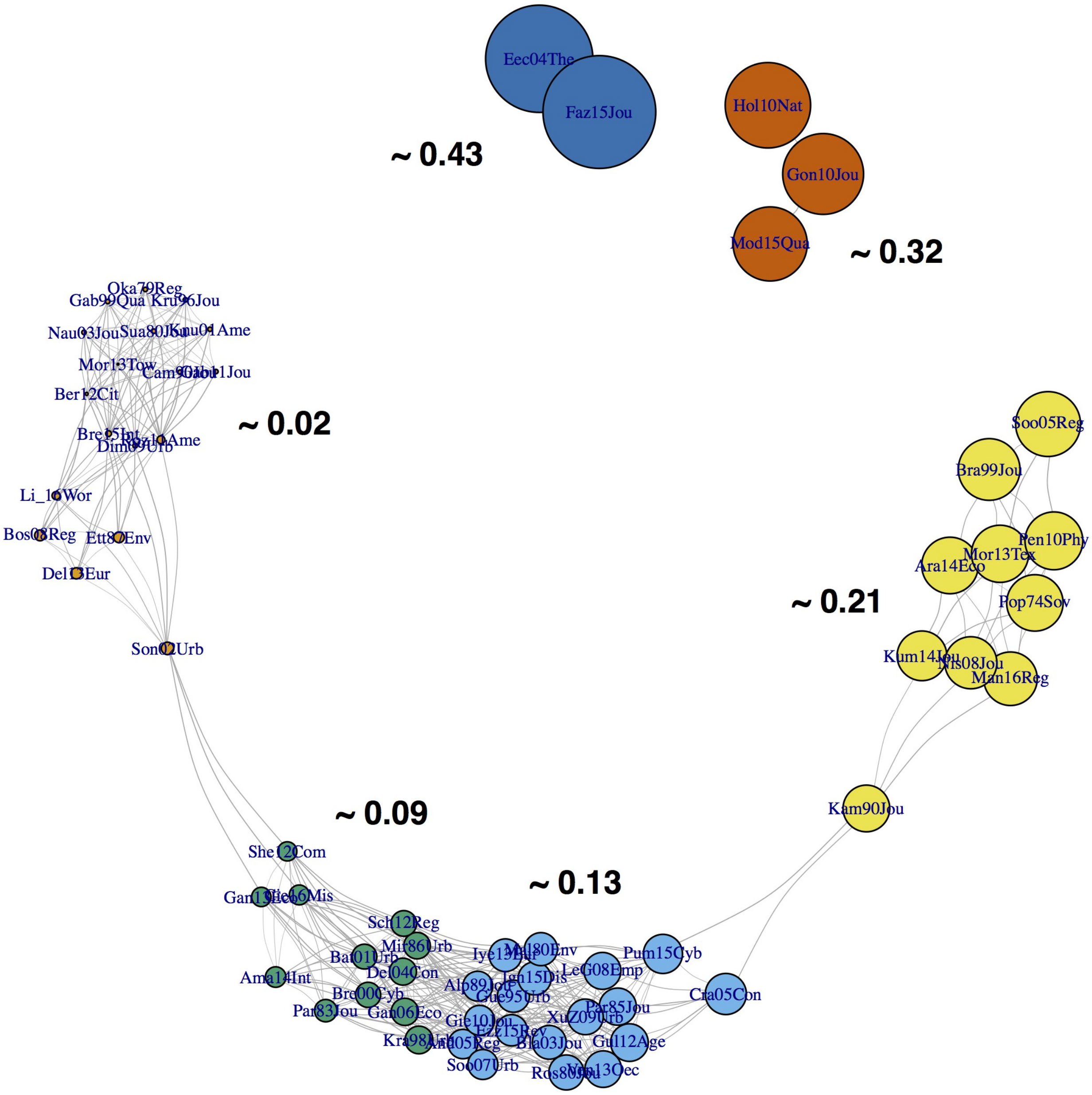

** the size of nodes reflects the standard deviation of estimates reported in the article and the numbers in black correspond to the average standard deviation for the community. cf. Table S0 in supplement for a lookup table between article ID and bibliographic references.*

Finally, we constructed a “n alpha” network to control for the number of estimates reported (especially when modelling their dispersion). We computed the number alpha estimates $n_i$ per study and the average value of alpha estimates $\bar{n}$ in the entire sample. We then computed a distance $dn_{ij}$ between studies as follows:

$$Dn_{ij} = | n_i - n_j | / \bar{n} \quad , \text{with } i \neq j$$

The smaller this distance the more studies i and j report a similar number of Zipf estimates. To transform this distance into a similarity, we simply multiply $d\bar{a}_{ij}$ by -1. The network emerging from this similarity is shown in suppementary figure S3.

**3.5. Modelling dyad similarities**

We run two series of models, one aimed at “explaining” the similarity in mean alpha values reported between corpus articles, and one aimed at explaining their similarity in alpha dispersion. “Explaning” variables for each series of models are similarity measures of the “wording”, “citation”, “disciplinary”, “country”, “decades”, “city definition” and “n alpha” networks. All variables were centered and scaled prior to modelling. We estimate the coefficients b and their residuals $e_{ij}$ by running step-wise OLS regressions.

$$MeanAlpha_{ij} = b_1\, Wording_{ij} + b_2\, Citation_{ij} + b_3\, Discipline_{ij} + b_4\, nAlpha_{ij} + b_4\, Country_{ij} + b_5\, Decade_{ij} + b_6\, CityDef_{ij} + e_{ij}$$

$$sdAlpha_{ij} = b_1\, Wording_{ij} + b_2\, Citation_{ij} + b_3\, Discipline_{ij} + b_4\, nAlpha_{ij} + b_4\, Country_{ij} + b_5\, Decade_{ij} + b_6\, CityDef_{ij} + e_{ij}$$

with i≠j, i and j being articles from the corpus.

We also look at interactions between control variables to identify studies which have studied similar national systems in the same decade and under the same definition of cities. These studies should report the most similar rank-size estimations.

## 4. Results

The results of the regressions are reporting in tables 2 and 3. Regarding the similarity in mean alpha reported in the corpus (table 2), we find a confirmation of two of our three initial hypotheses. Although the $R^2$ are low, the similarity in mean alpha varies positively and significantly with both the similarity in wording and the similarity in citations (models 1, 2 and 5). This means that articles written with a similar set of words and references tend to report similar values of Zipf estimates on average. This interesting feature persists (model 5) even when we account for the similarity in countries, decades and city definitions which the pair of corpus articles studies. As the wording network showed, this could result from a different setup from which the estimation originates. In some articles, the goal is to validate a “law” and the adequacy of one case to the “model”. It is thus more probable that such studies report estimates centered around 1, as in the strict version of Zipf's law. On the other hand, articles citing the same pool of references can exhibit a similar interest in validating or challenging the law. The evidence for the similarity in disciplinary references is more mixed, since the significant effect of the simple model 3 disappears when other variables and controls are accounted for. In terms of controls, we find that articles reporting a similar number of estimates tend to differ in mean alpha. This can be the effect of sensitivity analysis studies which explore the effect of threshold values or other specification criteria: they generally report a high number of estimates but their dispersion is such that the average value varies a lot. As expected, studies dedicated to the same set of countries tend to report similar values of estimates on average, however the opposite is true for time periods. The effect of similar city definitions chosen to analyse size distributions was not found significant by itself but appeared positive in conjunction with a similarity in the set of countries and with a similarity in the set of decades studied, as expected.

**Table 2. OLS regression of the similarity in average value of alpha reported in the corpus:**

| | *Dependent variable:* | | | | |
|---|---|---|---|---|---|
| | similarity in meanAlpha | | | | |
| Similarity in ... | (1) | (2) | (3) | (4) | (5) |
| wording | 0.048** | | | | 0.050** |
| | (0.022) | | | | (0.023) |
| citation | | 0.062*** | | | 0.069*** |
| | | (0.022) | | | (0.024) |
| discipline | | | 0.043* | | 0.015 |
| | | | (0.022) | | (0.024) |
| nAlpha | | | | -0.053** | -0.046** |
| | | | | (0.022) | (0.022) |
| country | | | | 0.063*** | 0.063*** |
| | | | | (0.023) | (0.023) |
| decade | | | | -0.100*** | -0.123*** |
| | | | | (0.022) | (0.023) |
| cityDef | | | | -0.001 | -0.015 |
| | | | | (0.022) | (0.023) |
| country:decade | | | | -0.014 | -0.013 |
| | | | | (0.021) | (0.021) |
| country:cityDef | | | | 0.049** | 0.053** |
| | | | | (0.022) | (0.022) |
| decade:cityDef | | | | 0.037* | 0.038* |
| | | | | (0.022) | (0.022) |
| country:decade:cityDef | | | | -0.017 | -0.014 |
| | | | | (0.021) | (0.021) |
| Constant | 0.015 | 0.015 | 0.015 | 0.013 | 0.012 |
| | (0.022) | (0.022) | (0.022) | (0.022) | (0.022) |
| Observations | 2,016 | 2,016 | 2,016 | 2,016 | 2,016 |
| $R^2$ | 0.002 | 0.004 | 0.002 | 0.021 | 0.030 |
| *Note:* | | | | | *p<0.1; **p<0.05; ***p<0.01 |

The distribution of positive residuals (figure S4) shows pairs with higher similarity than expected by the model. No obvious pattern seem to govern the association between such pairs, where luck might play a role. However, negative residuals are driven by three studies whose average estimate value differ from that of all others: Luckstead & Devadoss (2014), Le Gallo & Chasco (2008) and Popov (1974). They report an average value of alpha respectively of 1.91, 1.73 and 1.45. Those are very far away from the expected linear exponent of Zipf's law, which might suggest that considering them as outlyers for a subsequent meta analysis might be a perspective.

**Table 3. OLS regression of the similarity in standard deviation of alpha reported:**

| | *Dependent variable:* | | | | | |
|---|---|---|---|---|---|---|
| | Similarity in sdAlpha | | | | | |
| Similarity in ... | (1) | (2) | (3) | (4) | (5) | (6) |
| wording | 0.112*** | | | | | 0.116*** |
| | (0.022) | | | | | (0.023) |
| citation | | -0.013 | | | | -0.006 |
| | | (0.022) | | | | (0.024) |
| discipline | | | -0.004 | | | -0.009 |
| | | | (0.022) | | | (0.024) |
| nAlpha | | | | 0.134*** | | 0.133*** |
| | | | | (0.022) | | (0.022) |
| country | | | | | -0.096*** | -0.078*** |
| | | | | | (0.023) | (0.022) |
| decade | | | | | -0.077*** | -0.076*** |
| | | | | | (0.022) | (0.023) |
| cityDef | | | | | 0.088*** | 0.083*** |
| | | | | | (0.022) | (0.022) |
| country:decade | | | | | 0.041* | 0.042** |
| | | | | | (0.021) | (0.021) |
| country:cityDef | | | | | 0.050** | 0.051** |
| | | | | | (0.022) | (0.022) |
| decade:cityDef | | | | | 0.050** | 0.051** |
| | | | | | (0.022) | (0.022) |
| country:decade:cityDef | | | | | -0.023 | -0.016 |
| | | | | | (0.021) | (0.021) |
| Constant | -0.009 | -0.009 | -0.009 | -0.008 | -0.014 | -0.013 |
| | (0.022) | (0.022) | (0.022) | (0.022) | (0.022) | (0.022) |
| Observations | 2,016 | 2,016 | 2,016 | 2,016 | 2,016 | 2,016 |
| $R^2$ | 0.012 | 0.0002 | 0.00002 | 0.018 | 0.027 | 0.055 |

Regarding the similarity in dispersion (table 3), we find that only one of our main hypotheses is verified: the more articles are written with similar words, the more similar they are in terms of standard deviation of alphas reported (models 1 and 6). Again, some articles are similar in their attempts at verifying the "law": they are written with mathematical language and tend to report few estimates close in value. Other articles have the goal of exploring the national variation of city size distributions or their sensitivity to technical specifications: they use words like "countries", "spatial" and "comparison" and tend to report a very dispersed set of results. We do not find any significant evidence of covariation between the similarity in bibliography and disciplines cited and the similarity in alpha dispersion. However, the number of estimates is shown to positively influence the similarity in dispersion, since more estimates tends to increase the dispersion on average. Studies which use similar city definitions tend to report similar dispersion. Finally, although the similarity in countries and decades studied is negatively associated with a similarity in dispersion per se, they are positively associated when in interaction with one another and with city definition (model 6). The distribution of residuals (figure S5) exhibits the same properties as that of the previous model: elective similarity between more or less isolated pairs of studies and polarised dissimilarity with a couple of articles, including Luckstead & Devadoss (2014).

## 5. Discussion & Conclusion

In this article, we have looked at the empirical literature on Zipfs law for cities from a network perspective. As a complement to previous meta-analyses, the present approach has shed light on the scientific text and context mobilized to report on city size distributions. As in Raimbault et al., 2019, it has used textual analysis and citation networks to reflect various proximities between articles of the corpus. The analysis of each network had produced insight in the wording, reference framework and disciplinary heritage demonstrated by the empirical literature on Zipf's law for cities. Their use as explaining variables of a model of the similarity in the distribution of estimates reported has shown that wording is important in both cases, whereas similar citation patterns mostly impact the average value of Zipf's estimate reported.

The contribution of this paper to meta-analyses has been two-fold. Firstly, using the citation networks of studies included in a meta-analysis has allowed us to identified gaps in the corpus and potentially overlooked articles. These have appeared when looking at the most cited external references. In our case, the article of Eaton & Eckstein (1997) for example is one of the most externally cited reference to report empirical estimations of Zipf's law. It was initially rejected from the corpus (Cottineau, 2017) because the estimation included instruments. The present analysis suggests that relaxing this criterion could allow its inclusion as a major reference in the field. Symmetrically, the analysis of model residuals has shown that some very atypical studies drive a large share of the difference in mean values and dispersion used in the meta-analysis, suggesting that removing them as outlyers could provide clearer results. Secondly, the data and code of the present study has been made open on Github, including an R notebook with all visualisations, in order to be reused by the community[8].

Although this article does not close the debate on city size distribution, it has tried to reveal a newer aspect of a literature in rapid development: the fact that it mixes studies of very different aims and methods, potentially characterised by reporting biases. What seems quite obvious from the corpus is also the fact that Zipf's law estimation is a large field where many authors contribute at one point of their scientific career in urban studies, economics or physics, but mostly is not a dominant object of individual research *per se*. A further point of inquiry in the reflexive meta-analysis could thus be to trace various authors' contribution to the empirical Zipf literature as part of their own scientific topic trajectory (Zeng et al., 2019). However, it is not obvious at this point to which extend it would help provide guidelines for rigorous analysis of city size distribution.

8 http://clementinecttn.github.io/MetaZipf/metametazipf_notebook.nb.html

# Supplementary Material

## Table S0. Correspondence between corpus articles and identifiers (reading help for graphs).

| REFID | AUTHOR | YEAR | JOURNAL |
|---|---|---|---|
| Alp89Jou | Alperovich G. | 1989 | Journal of Urban Economics |
| Ama14Int | Amalraj V. C. & Subbarayan A. | 2014 | International Journal of Pure and Applied Mathematics |
| And05Reg | Anderson G. & Ge Y. | 2005 | Regional Science and Urban Economics |
| Ara14Eco | Aragon J. A. O. & Queiroz V. | 2014 | EconomiA |
| Arr12Reg | Arribas-Bel D. et al. | 2012 | Region et Developpement |
| Bat01Urb | Batty M. | 2001 | Urban Studies |
| Ber12Cit | Berry B. & Okulicz-Kozaryn A. | 2012 | Cities |
| Bla03Jou | Black D. & Henderson V. | 2003 | Journal of Economic Geography |
| Bos08Reg | Bosker al. | 2008 | Regional Science and Urban Economics |
| Bra99Jou | Brakman S. et al. | 1999 | Journal of Regional Science |
| Bre00Cyb | Bretagnolle A. et al. | 2000 | Cybergeo |
| Bre15Int | Bretagnolle A. & Delisle F. | 2015 | International Journal of Geographical Information Science |
| Cam90Jou | Cameron T. A. | 1990 | Journal of Urban Economics |
| Cie16Mis | Cieslik A. & Teresnski J. | 2016 | Miscellanea Geographica |
| Cra05Con | Crampton G. | 2005 | Congress of the European Regional Science Association |
| Del04Con | Delgado A. P. & Godinho I. M. | 2004 | Congress of the European Regional Science Association |
| Del13Eur | Deliktas E. et al. | 2013 | European Planning Studies |
| Dim09Urb | Dimou M. & Schaffar A. | 2009 | Urban Studies |
| Eec04The | Eeckhout J. | 2004 | The American Economic Review |
| Ett87Env | Ettlinger N. & Archer J. C. | 1987 | Environment and Planning A |
| Ezz15Rev | Ezzahid E. & ElHamdani O. | 2015 | Review of Urban & Regional Development Studies |
| Faz15Jou | Fazio G. & Modica M. | 2015 | Journal of Regional Science |
| Gab11Jou | Gabaix X. & Ibragimov R. | 2011 | Journal of Business & Economic Statistics |
| Gab99Qua | Gabaix X. | 1999 | Quarterly Journal of Economics |
| Gan06Eco | Gan L. et al | 2006 | Economics Letters |
| Gan13Eco | Gangopadhyay K. & Basu B. | 2013 | Econophysics of Systemic Risk and Network Dynamics |
| Gie10Jou | Giesen K. & Sudekum J. | 2010 | Journal of Economic Geography |
| Gon10Jou | Gonzalez-Val R. | 2010 | Journal of Regional Science |
| Gue95Urb | Guerin-Pace F. | 1995 | Urban Studies |
| Gul12Age | Gulden T. R. & Hammond R. A. | 2012 | Agent-based models of geographical systems |
| Hol10Nat | Holmes T. J. & Lee S. | 2010 | National Bureau of Economic Research |
| Ign15Dis | Ignazzi A. | 2015 | Cybergeo |
| Iye13Eur | Iyer S. D. | 2013 | Eurasian Geography and Economics |
| Kam90Jou | Kamecke U. | 1990 | Journal of Urban Economics |
| Knu01Ame | Knudsen T. | 2001 | American Journal of Economics and Sociology |
| Kra98Urb | Krakover S. | 1998 | Urban Studies |
| Kru96Jou | Krugman P. | 1996 | Journal of the Japanese and International Economies |
| Kum14Jou | Kumar G. & Subbarayan A. | 2014 | Journal of Mathematics and Statistics |
| LeG08Emp | LeGallo J. & Chasco C. | 2008 | Empirical Economics |
| Li_16Wor | Li C. & Gibson J. | 2016 | Working Paper in Economics 09/16 |
| Luc14Eco | Luckstead J. & Devadoss S. | 2014 | Economics Letters |
| Mal80Env | Malecki E. J. | 1980 | Environment and Planning A |
| Man16Reg | Manaeva I. & Rastvortseva S. | 2016 | Regional Science Inquiry |
| Mir86Urb | Mirucki J. | 1986 | Urban Studies |
| Mod15Qua | Modica M. et al. | 2015 | Quaderni - Working Paper DSE |
| Mor13Tex | Moro S. & Santos R. | 2013 | Textos para Discussion Cedeplar-UFMG |
| Mor13Tow | Morudu H. & du Plessis D. | 2013 | Town and Regional Planning |
| Nau03Jou | Naude W.A. & Krugell W.F. | 2003 | Journal of African Economies |
| Nis08Jou | Nishiyama Y. et al. | 2008 | Journal of Regional Science |
| Oka79Reg | Okabe A. | 1979 | Regional Science and Urban Economics |
| Par83Jou | Parr J. B. & Jones | 1983 | Journal of Regional Science |
| Par85Jou | Parr J. B. | 1985 | Journal of Urban Economics |
| Pen10Phy | Peng G. | 2010 | Physica A: Statistical Mechanics and its Applications |
| Pop74Sov | Popov V. R. | 1974 | Soviet Geography |
| Pum15Cyb | Pumain D. et al. | 2015 | Cybergeo |
| Ros80Jou | Rosen K. T. & Resnick M. | 1980 | Journal of Urban Economics |
| Roz11Ame | Rozenfeld al. | 2011 | American Economic Review |
| Sch12Reg | Schaffar A. & Dimou M. | 2012 | Regional Studies |
| She12Com | Shepotylo O. | 2012 | Comparative Economic Studies |
| Son02Urb | Song S. & Zhang K. H. | 2002 | Urban Studies |
| Soo05Reg | Soo K. T. | 2005 | Regional Science and Urban Economics |
| Soo07Urb | Soo K. T. | 2007 | Urban Studies |
| Sua80Jou | Suarez-villa L. | 1980 | Journal of Regional Science |
| Ven13Oec | Veneri P. | 2013 | Oecd Publishing |
| XuZ09Urb | Xu Z. & Zhu N. | 2009 | Urban Studies |
| Ziq16Asi | Ziqin W. | 2016 | Asian Journal of Social Science Studies |

**Table S1. Correspondence between ad-hoc fields and Scimago classification of the most externally cited journals (at least five citations from the corpus).**

| Journal | Cites | Group | Scopus Subject Area & category 1 | Scopus Subject Area & category 2 |
|---|---|---|---|---|
| **Journal Regional Science** | 30 | REG | Environmental Science | Development |
| **Urban Studies** | 26 | REG | Environmental Science | Urban Studies |
| **Journal Urban Economics** | 25 | ECO | Economics and Econometrics | Urban Studies |
| **American Economic Review** | 16 | ECO | Economics and Econometrics | x |
| **Economic Development Cultural Change** | 15 | ECO | Economics and Econometrics | Development |
| **Econometrica** | 14 | ECO | Economics and Econometrics | x |
| **Environment Planning A** | 14 | GEO | Geography, Planning and Development | Environmental Science |
| **Quarterly Journal Economics** | 14 | ECO | Economics and Econometrics | x |
| **Regional Science Urban Economics** | 14 | REG | Urban Studies | Economics and Econometrics |
| **International Regional Science Review** | 11 | REG | Environmental Science (miscellaneous) | Social Sciences |
| **Physica A** | 11 | PHY | Condensed Matter Physics | Statistics and Probability |
| **Papers in Regional Science** | 15 | REG | Environmental Science | Geography, Planning and Development |
| **Annals AAG** | 9 | GEO | Geography, Planning and Development | Earth-Surface Processes |
| **Geographical Analysis** | 9 | GEO | Geography, Planning and Development | Earth-Surface Processes |
| **Journal Political Economy** | 9 | ECO | Economics and Econometrics | x |
| **Journal Econometrics** | 8 | ECO | Economics and Econometrics | Applied Mathematics |
| **Journal Economic Geography** | 8 | ECO | Economics and Econometrics | Geography, Planning and Development |
| **NBER** | 8 | ECO | x | x |
| **Annals Regional Science** | 6 | REG | Environmental Science | Social Sciences |
| **CEPR** | 6 | ECO | x | x |
| **Dissertation** | 6 | OTHER | x | x |
| **Geographical Review** | 6 | GEO | Geography, Planning and Development | Earth-Surface Processes |
| **Journal American Statistical Association** | 6 | STAT | Statistics, Probability and uncertainty | Statistics and Probability |
| **Region Development** | 6 | REG | x | x |
| **Journal Royal Statistical Society** | 5 | STAT | Statistics and Probability | Economics and Econometrics |
| **Physical Review E** | 5 | PHY | Condensed Matter Physics | Statistical and Nonlinear Physics |
| **Professional Geographer** | 5 | GEO | Geography, Planning and Development | Earth-Surface Processes |
| **Regional Studies** | 5 | REG | Environmental Science | Social Sciences |
| **Review Economic Studies** | 5 | ECO | Economics and Econometrics | x |
| **World Development** | 5 | OTHER | Development | Sociology and Political Science |
| **WorldBank** | 5 | STAT | x | x |

**Figure S1. Distribution and density of corpus articles over the years, according to their citation of Zipf's works or not.**

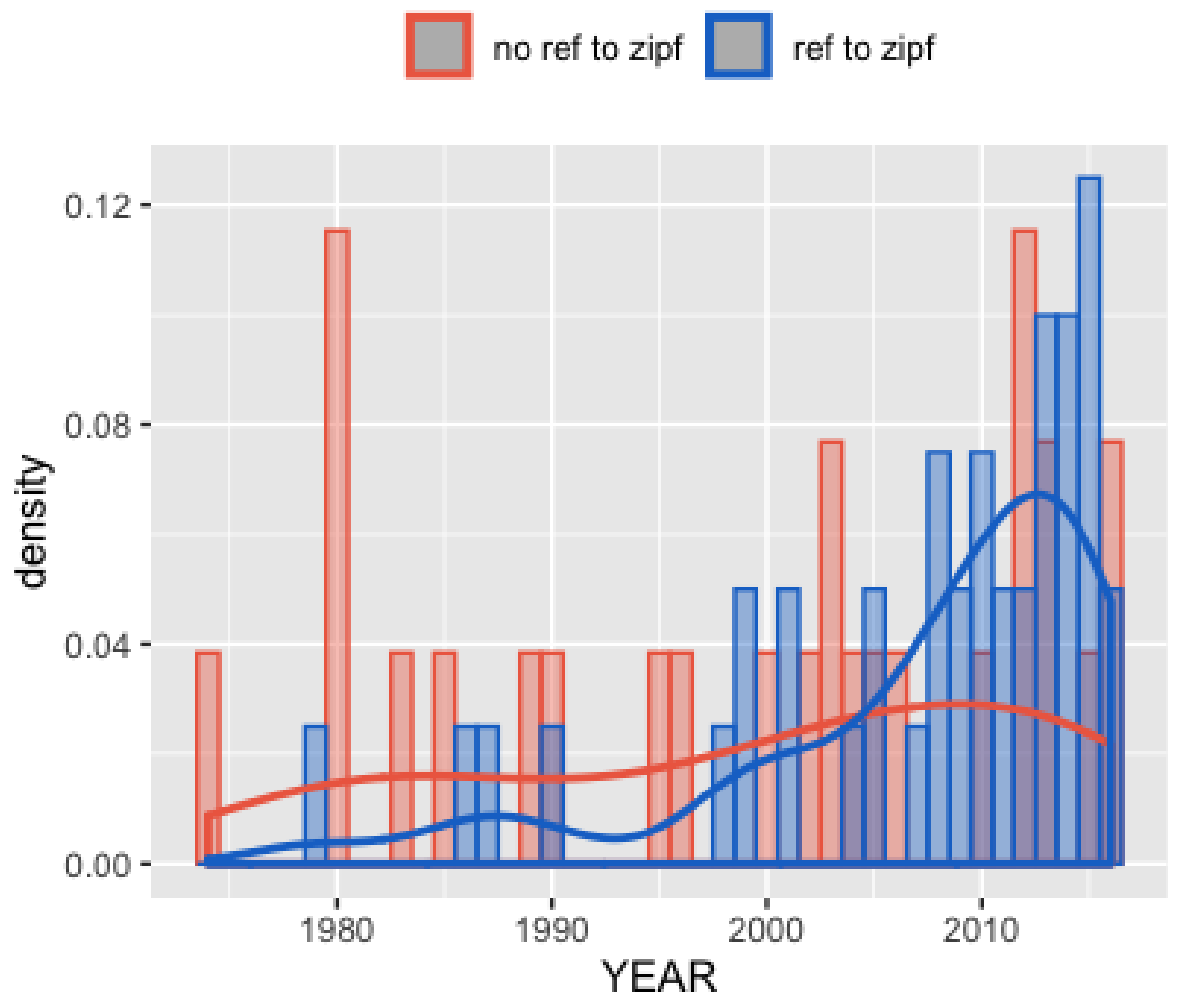

**Figure S2. Similarity network of corpus articles by the common countries they reported alpha on (cut-off 0.25).** The size of vertices represent the number of countries they report estimates for.

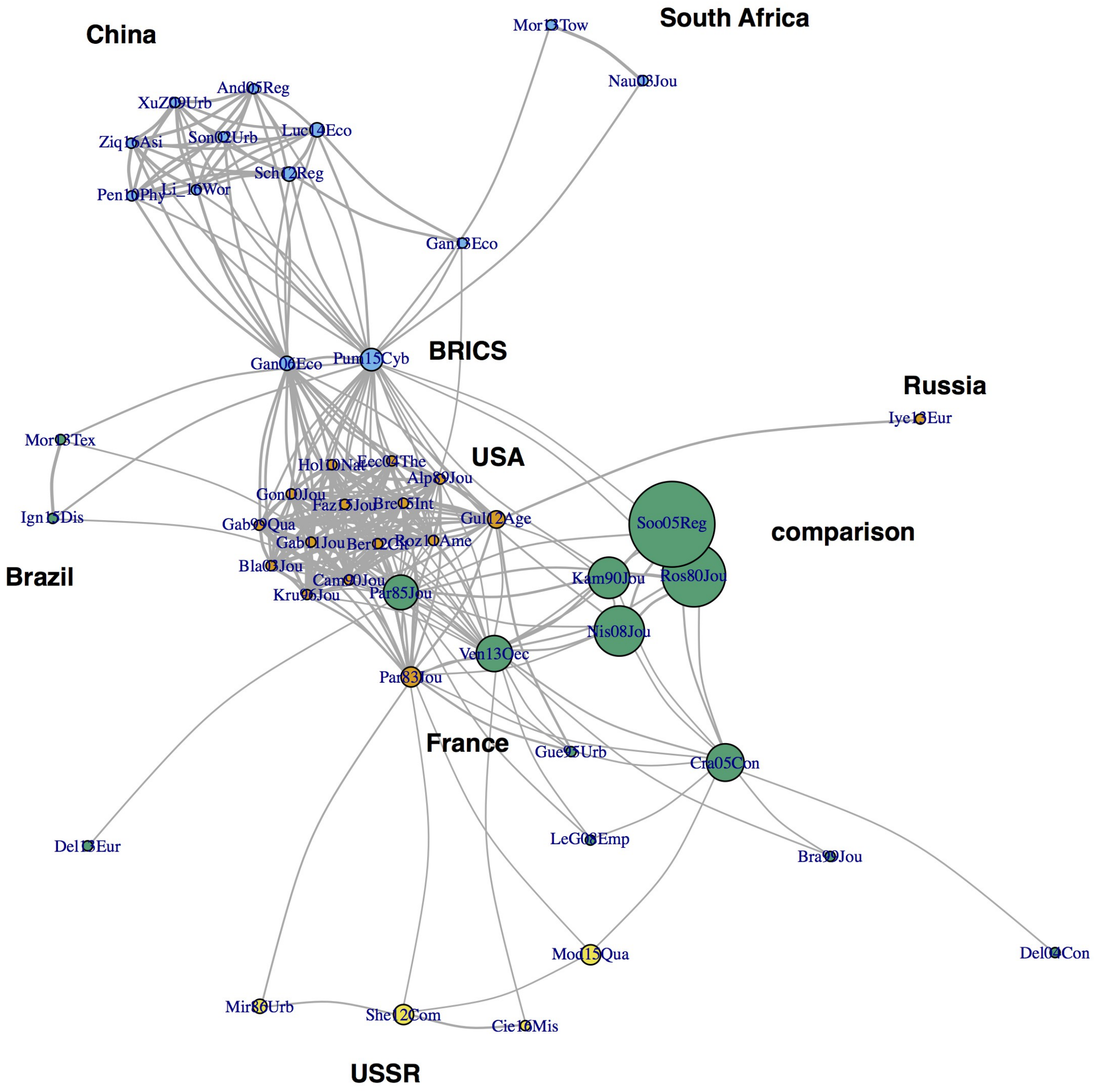

**Figure S3. Similarity network of corpus articles by the common decades they reported alpha on (cut-off at 0.65).**

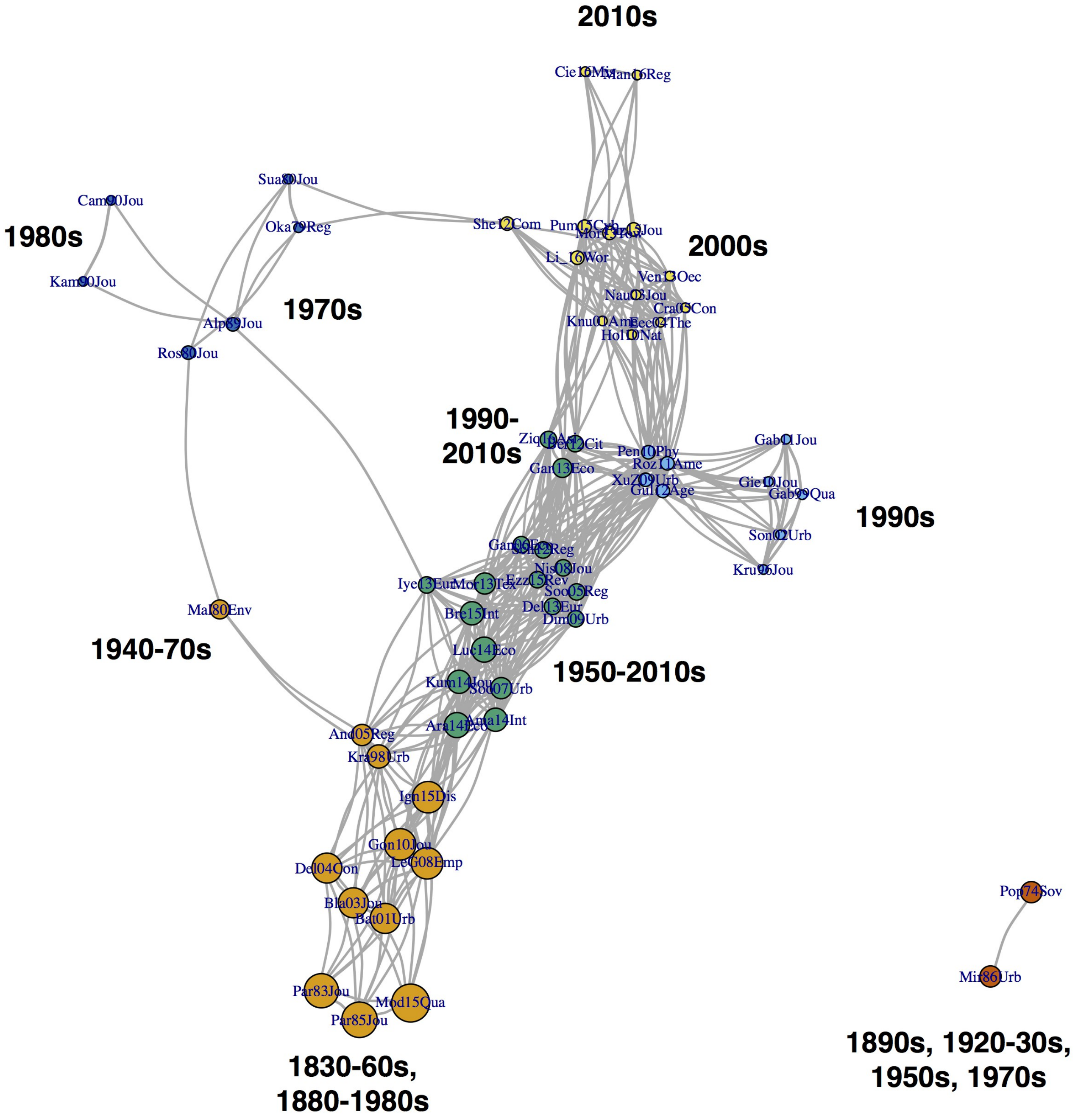

**Figure S4. Similarity network of corpus articles by the common city definitions they reported alpha on (cut-off at 0.1).**

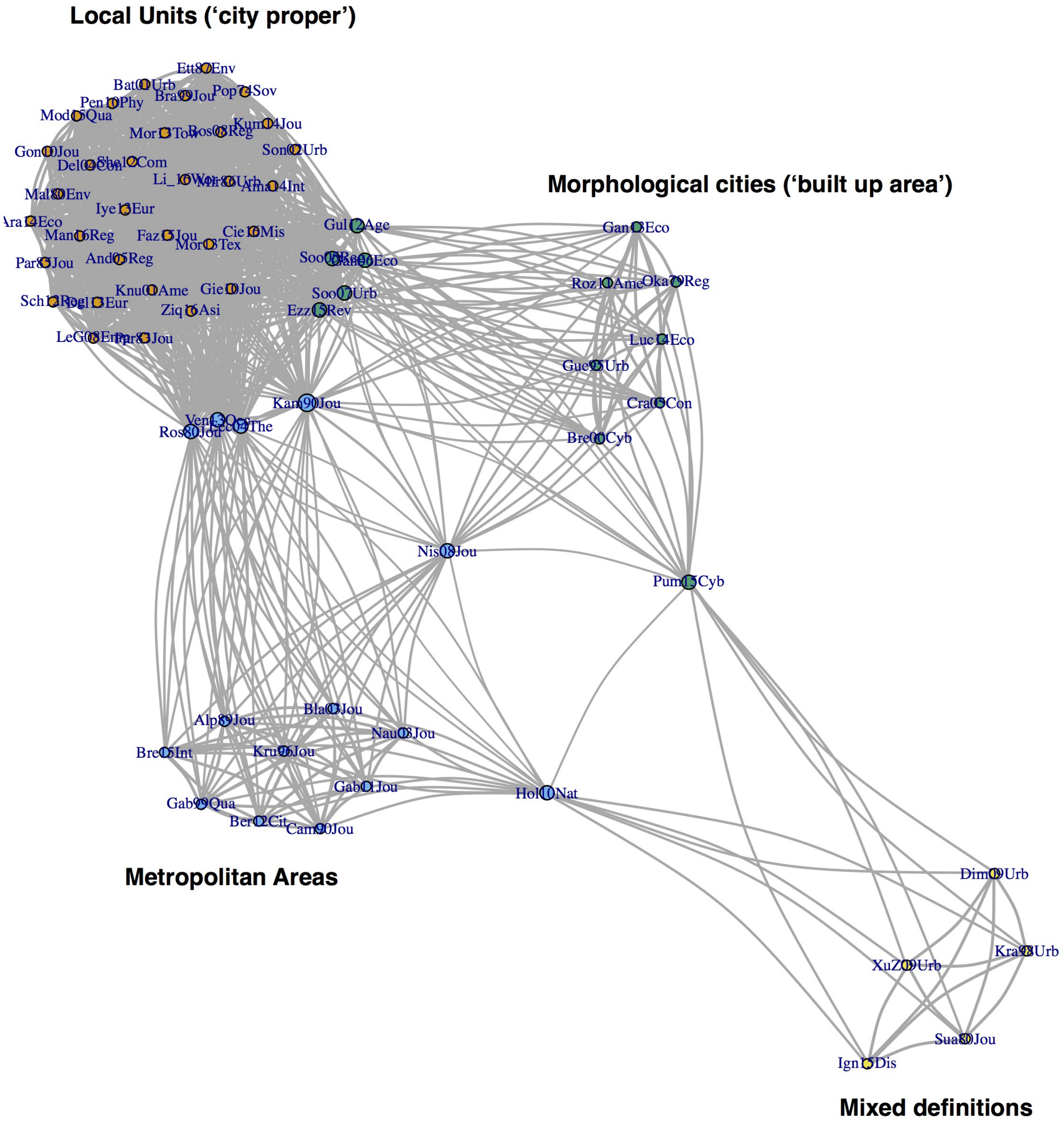

**Figure S3. Similarity network of corpus articles by number of estimates reported (cut-off at -0.1).**

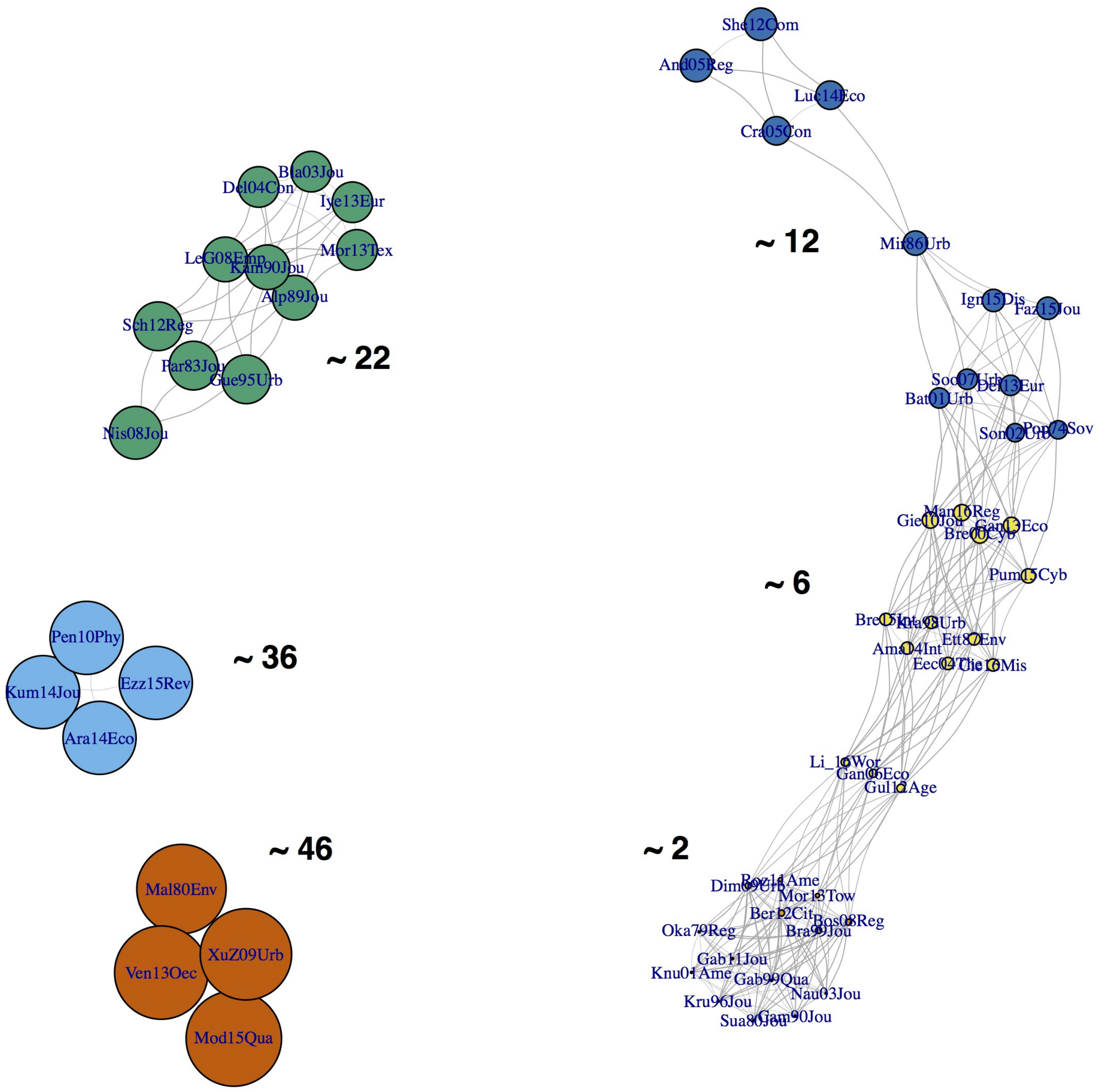

** the size of nodes reflects the number of estimates reported in the article and the numbers in black correspond to the average number of estimate per articles reported for the community.*

**Figure S4. Residuals of the model of similarity in mean alpha. Left: Most positive residuals (over 1.1). Right: Most negative residuals (under -2).**

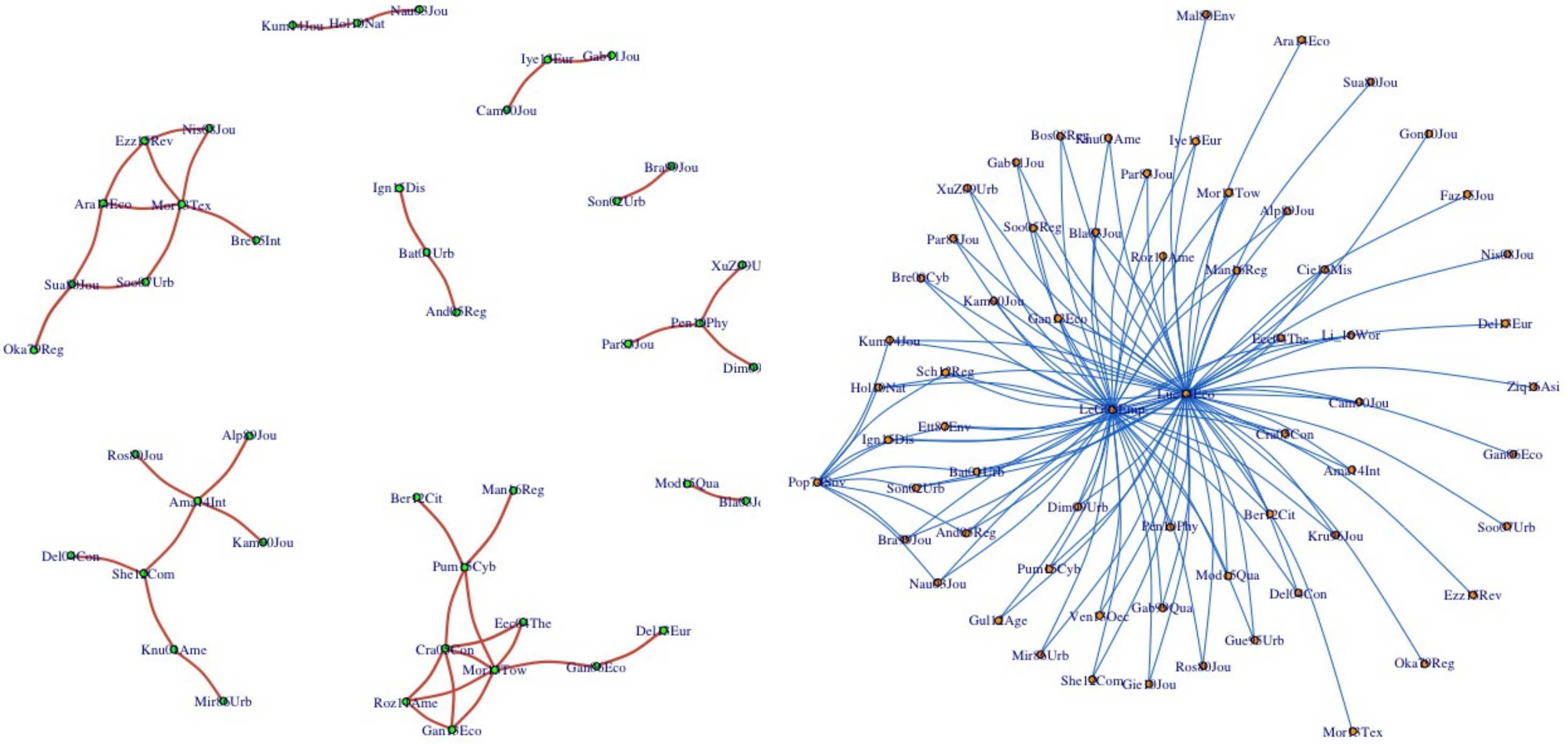

**Figure S5. Residuals of the model of similarity in alpha dispersion. Left: Most positive residuals (over 1). Right: Most negative residuals (under -1.5).**

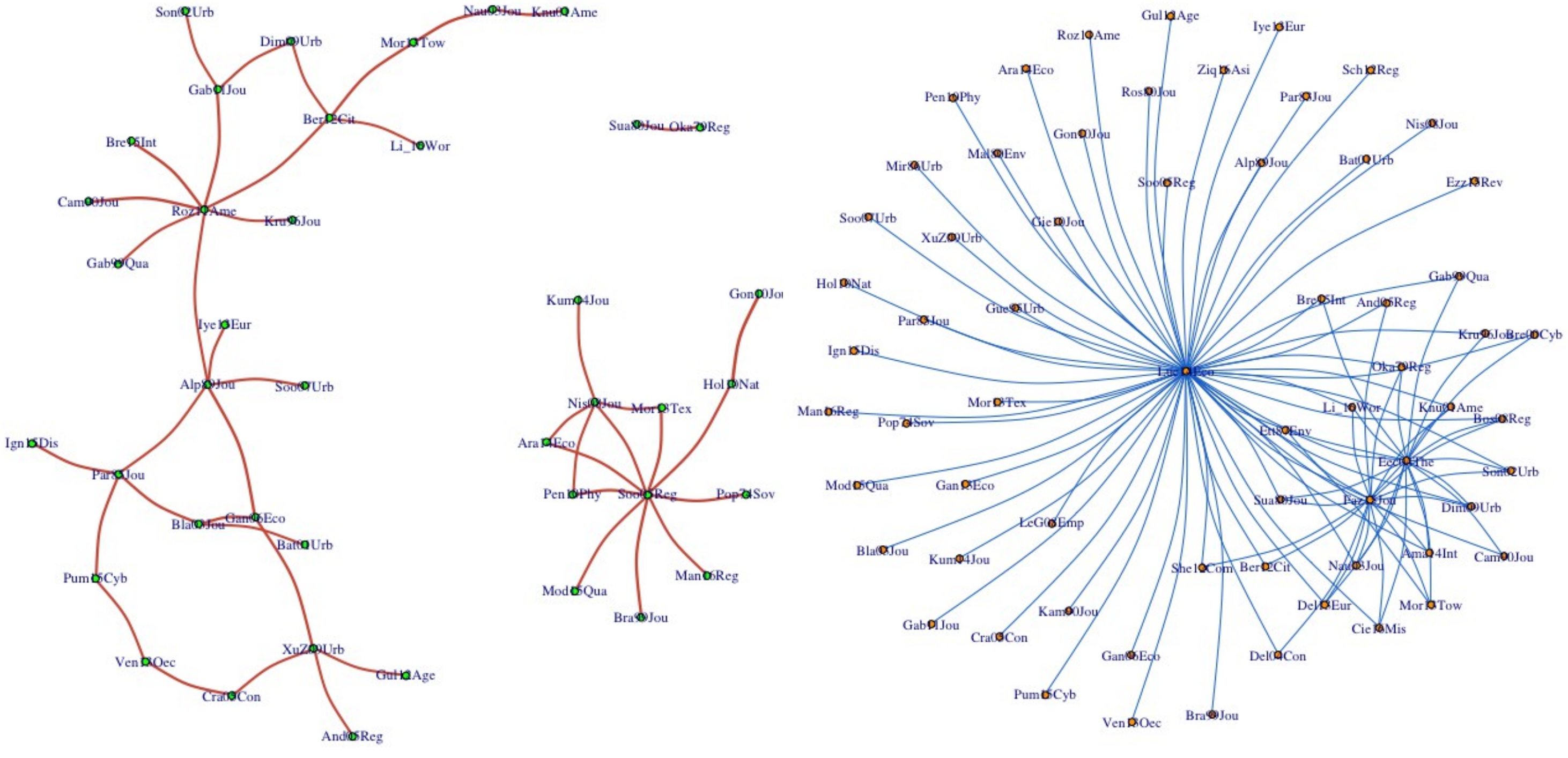